\documentclass{ngc4861ULX2} 

\usepackage{natbib}
\usepackage{hyperref}
\usepackage[T1]{fontenc}
\usepackage{ae,aecompl}
\usepackage{rotating}
\usepackage{color}
\usepackage{graphicx}
\usepackage{appendix}
\usepackage{longtable}

\usepackage{txfonts}
\begin{document} 

\titlerunning{A candidate cyclotron line in NGC 4861 X--2}

\authorrunning{S. Allak et al.} 
 \title{A candidate cyclotron line at 1.89 keV in the ultraluminous X-ray source NGC 4861 X--2}

 \author{Sinan Allak\inst{1}, Lorenzo Ducci\inst{1}, Valery F. Suleimanov\inst{1}, Andrea Santangelo\inst{1}, Aysun Akyuz\inst{2}, Santina Piraino\inst{1}, Faruk Soydugan\inst{3}, Amar Deo Chandra\inst{4}, Wei Yu\inst{1}}

\institute{Institut für Astronomie und Astrophysik, Sand 1, 72076 
Tübingen, Germany\\
\email{sinan.allak@uni-tuebingen.de}
\and
Department of Physics, University of Çukurova, 01330, Adana, Türkiye
\and
Department of Physics, University of Çanakkale Onsekiz Mart, 17100, Çanakkale, Türkiye
\and
Aryabhatta Research Institute of Observational Sciences, Manora Peak, Nainital, Uttarakhand, 263001, India
}

\abstract{In this Letter, we report the detection of an absorption-like feature at $\sim 1.89$ keV in \textit{Chandra}/ACIS spectra of ultraluminous X-ray source NGC\,4861 X--2, based on the deepest observation (ObsID 20992; $\sim58$ ks). The feature is consistently recovered across independent continuum models and significantly improves the fit statistics. Monte Carlo simulations yield a detection significance of $\sim3.5$--$4.1\sigma$, depending on the adopted continuum, and a blind line scan reveals a single, localized peak at the same energy. The observed properties are consistent with a proton cyclotron resonant scattering feature (CRSF), implying a magnetic field strength of $B\sim(3$--$4)\times10^{14}$ G. The spectrum is well described by a multicolor disk blackbody (\textit{diskbb}) with $kT_{\rm in}\sim 0.8$ keV or a strongly curved continuum with a low cutoff (\textit{cutoffpl}) energy ($E_{\rm cut}\sim1.3$ keV). The source shows variability confined to the soft X-ray band in the two \textit{Chandra} observations where the absorption-like feature is detected. In these observations, a candidate periodic signal at $P\approx7.4$ s is also detected, with a global significance of $\sim2.5\sigma$.}


 \maketitle
%
%-------------------------------------------------------------------

\section{Introduction}

Ultraluminous X-ray sources (ULXs) are off-nuclear point sources with apparent X-ray luminosities exceeding $2\times10^{39}$ erg s$^{-1}$, above the Eddington limit for typical stellar-mass compact objects \citep{2017ARA&A..55..303K,2021AstBu..76....6F,2023NewAR..9601672K,2023arXiv230200006P}. The currently favored interpretation is that most ULXs are powered by super-Eddington accretion onto stellar-mass black holes or neutron stars (NSs), where geometric beaming and radiation-driven outflows can significantly enhance the apparent luminosity \citep{2007MNRAS.377.1187P,2009MNRAS.393L..41K,2016Natur.533...64P,2018MNRAS.479.3978K}. The discovery of coherent X-ray pulsations in several ULXs has provided unambiguous evidence that a subset of the population hosts accreting NSs \citep{2014Natur.514..202B,2014Natur.514..198M,2017Sci...355..817I,2018MNRAS.476L..45C,2018ApJ...863....9W, 2020ApJ...895...60R, 2025ApJ...994L..38D}. These sources provide important laboratories for studying accretion physics in the presence of strong magnetic fields. Recent studies suggest that some ULX pulsars may host strong dipolar magnetic fields of order $\sim10^{13}$--$10^{14}$ G, potentially enabling super-Eddington accretion via magnetically channeled inflow (e.g., \citealt{2020ApJ...899...97E}). 

In this context, cyclotron resonant scattering features (CRSFs) provide the only direct observational probe of the magnetic field strength near the surface of accreting NSs \citep{1992herm.book.....M,2012MmSAI..83..230C,2019A&A...622A..61S}. These absorption-like features arise from resonant scattering of X-ray photons by charged particles quantized in Landau levels in strong magnetic fields \citep{2017A&A...601A..99S}. These features frequently exhibit harmonic structures and strong pulse-phase dependence, reflecting the complex geometry of the accretion column and magnetic field configuration \citep{2012MNRAS.420.2307M,2019A&A...622A..61S}. Candidate CRSFs have also been reported in extragalactic ULXs, including M51 ULX-8 \citep{2018NatAs...2..312B} and NGC 4656 ULX-1 \citep{2026arXiv260310331C}, extending CRSF phenomenology into the ultraluminous regime. CRSFs can arise from either electrons or protons and therefore probe different magnetic field regimes. Electron cyclotron lines are typically observed at tens of keV, implying magnetic field strengths of $\sim10^{12}$ G \citep{1978ApJ...219L.105T,2002ApJ...580..394C}. In contrast, proton cyclotron features are expected at much lower energies (typically $\sim 0.1$--a few keV) for magnetar-strength fields of $B\sim10^{14}$--$10^{15}$ G \citep{2006RPPh...69.2631H, 2013Natur.500..312T}. The detection of an isolated absorption feature at soft X-ray energies without harmonics is therefore commonly interpreted as indicative of a proton CRSF \citep{2010A&A...518A..24P}.

NGC\,4861 (Mrk\,59) is a Magellanic-type irregular galaxy at a distance of 9.95\,Mpc \citep{2013AJ....146...86T}. It hosts two ULXs, X--1 and X--2, identified with \textit{ROSAT} \citep{2002ApJS..143...25C,2005ApJS..157...59L}. Follow-up \textit{XMM--Newton} and \textit{Chandra} observations associate these sources with H\,II regions and massive stars, and in the \textit{Chandra} observation, a thermal disk model is preferred with a characteristic temperature of $kT \sim 0.80\,\mathrm{keV}$, and its luminosity lies in the range $(1{-}3)\times10^{39}\,\mathrm{erg\,s^{-1}}$ \citep{2014MNRAS.441.1841T,2021MNRAS.505..771O}. In this Letter, we report the detection of an absorption feature at $\sim 1.89$ keV in the \textit{Chandra}/ACIS-S spectrum of NGC\,4861 X--2. The feature is consistently recovered across different continuum models and is supported by Monte Carlo simulations and a blind line scan. Its properties favor a proton CRSF interpretation, implying a magnetic field strength in the magnetar regime ($B \sim 10^{14}$--$10^{15}$ G).

\section{Observation, data reduction and analysis} 

\subsection{Observation}

NGC 4861 was observed multiple times with \textit{Chandra} and \textit{XMM--Newton}. The available \textit{Chandra} dataset consists of one ACIS-I observation and four ACIS-S observations obtained between 2012 and 2018, while three \textit{XMM--Newton} observations were carried out in 2003. A summary of the observations used in this work is provided in Table~\ref{tab:obs_log}.

\subsection{Energy spectra}

The spectral analysis was performed using \textsc{XSPEC} \citep{Arnaud1996} over the 0.3--10 keV band for the deeper \textit{Chandra}/ACIS-S observation (ObsID 20992; $\sim$58 ks), employing a range of spectral models, including single-component continua such as \textit{diskbb}, a power law (pl), and a cutoff power law (\textit{cutoffpl}), as well as combinations of these components, together with photoelectric absorption modeled using \textit{tbabs}. The Galactic column density was fixed at $N_{\mathrm{H,Gal}} = 1.3 \times 10^{20}\,\mathrm{cm}^{-2}$ \citep{1990ARA&A..28..215D}, while any additional absorption was allowed to vary unless otherwise noted. Given the limited number of source counts, all spectral fitting was performed using C-statistics. The spectra were grouped using both 10 and 15 counts per bin in order to assess the robustness of the detected feature against spectral grouping. Unless otherwise stated, all parameter uncertainties are quoted at the 90\% confidence level. Among the models considered, the \textit{diskbb} and \textit{cutoffpl} models provided the best phenomenological descriptions of the spectrum and were therefore adopted for the subsequent analysis. The \textit{diskbb} model yields an inner-disk temperature of $kT_{\mathrm{in}} = 0.79^{+0.06}_{-0.05}$ keV. In this case, the intrinsic column density is not constrained by the data and was therefore fixed at $N_{\mathrm{H}} = 1.3 \times 10^{20}\,\mathrm{cm}^{-2}$. The \textit{cutoffpl} model gives a hard photon index of $\Gamma = 0.76^{+0.51}_{-0.54}$, together with a low cutoff energy of $E_{\mathrm{cut}} = 1.37^{+0.74}_{-0.37}$ keV, indicating strong spectral curvature already within the \textit{Chandra} band.

\begin{figure}
\resizebox{\hsize}{!}{\includegraphics{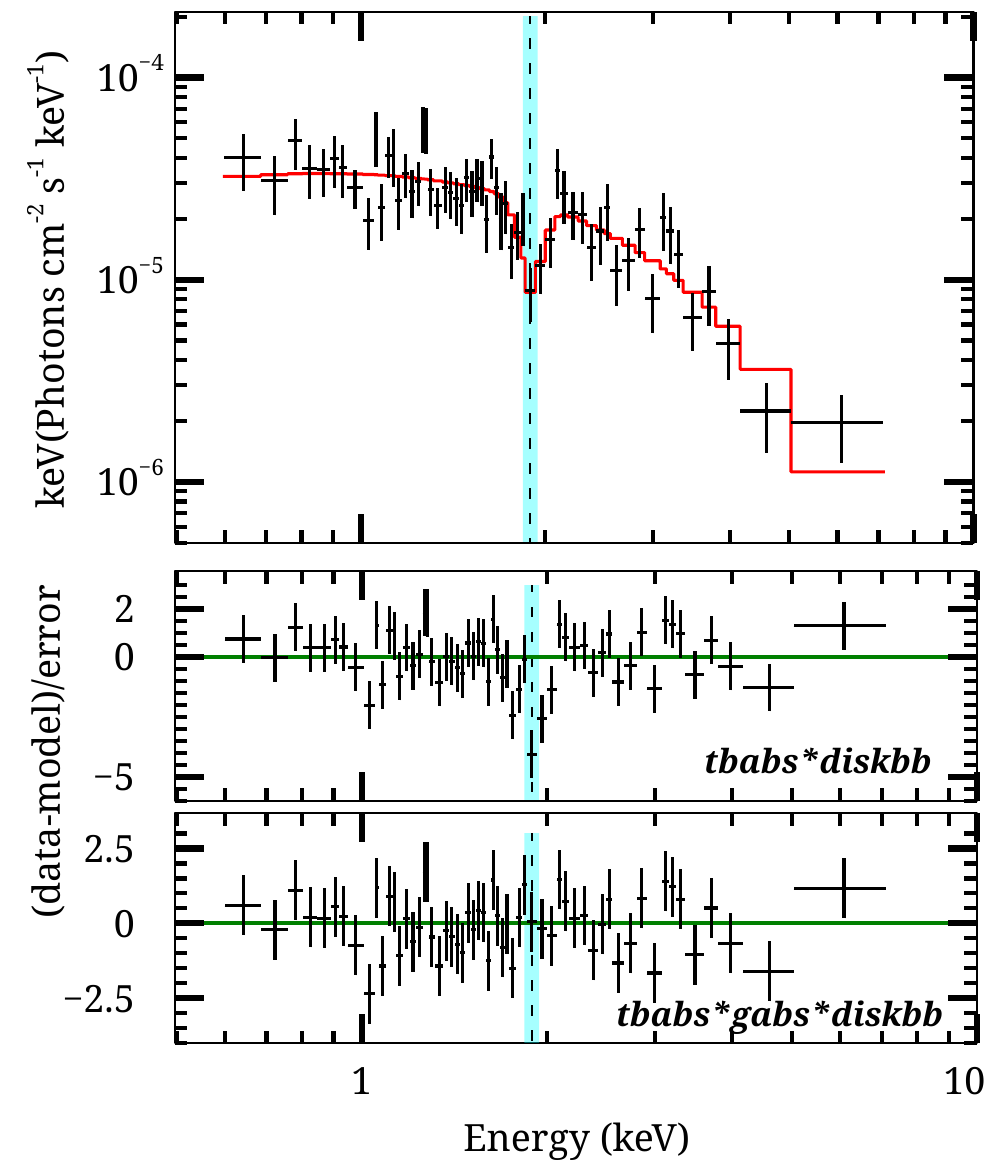}}
\caption{\textit{Chandra}/ACIS-S (ObsID 20992) energy spectrum of NGC\,4861 X--2. Top panel: spectrum fitted with the \textit{tbabs*gabs*diskbb} model (red line). Middle panel: residuals for the continuum-only model (\textit{tbabs*diskbb}), showing a clear absorption-like deficit at $\sim 1.89$ keV. Bottom panel: residuals after including the multiplicative Gaussian absorption component (\textit{gabs}). The shaded region marks the centroid energy of the absorption feature. }
\label{fig:spectrum}
\end{figure}

Although both continua provide statistically acceptable descriptions of the overall spectrum, they leave a localized negative residual at $\sim 1.89$ keV (Fig.~\ref{fig:spectrum}, middle panel). The residual appears at the same energy in both models, supporting its model-independent nature. We therefore added a multiplicative Gaussian absorption component, \textit{gabs}, to each continuum. The best-fit line parameters are consistent across different continuum models, with a centroid energy of $E_{\rm line}\sim1.89$ keV, a width of $\sigma \sim 0.08$ keV, and a line strength of $S_{\rm line} \sim 0.2$ keV. The addition of the absorption component leads to a significant improvement in the fit, yielding $\Delta C = 20.96$ and $\Delta C = 17.23$ for the spectra grouped to 10 counts per bin using the \textit{diskbb} and \textit{cutoffpl} models, respectively. Consistent improvements are also obtained for the spectra grouped to 15 counts per bin using C-statistics, with $\Delta C = 20.04$ and $\Delta C = 17.35$, respectively. As shown in the lower panel of Fig.~\ref{fig:spectrum}, including the absorption component removes the residual structure around 1.89 keV. This indicates that the feature is not driven by the choice of fitting statistic. We derived the statistical significance of the absorption feature using Monte Carlo simulations with the XSPEC \textit{simftest} routine (see Appendix~\ref{blind}). The feature is detected with significances of $\sim4.1\sigma$ and $\sim3.5\sigma$ for the \textit{diskbb} and \textit{cutoffpl} continua, respectively. The blind line scan also reveals a single localized peak at $\sim1.89$ keV (see Appendix~\ref{blind}), supporting the robustness of the feature against continuum uncertainties. These results indicate that the inclusion of the \textit{gabs} component provides a statistically significant improvement over the corresponding continuum-only models. Table~\ref{tab:spec_all} summarizes the best-fit spectral parameters for the adopted continuum and line models.

\begin{table}[t]
\centering
\caption{Best-fit spectral models and parameters for X--2.}
\label{tab:spec_all}
\begin{tabular}{lcccc}
\hline
Par/models & A & B & C & D \\
\hline
$kT_{\rm in}$ & $0.79^{+0.06}_{-0.05}$ & $0.80^{+0.06}_{-0.05}$ & -- & -- \\
$\Gamma$ & -- & -- & $0.76^{+0.51}_{-0.54}$ & $0.23^{+0.64}_{-0.63}$ \\
$E_{\rm cut}$ & -- & -- & $1.37^{+0.74}_{-0.37}$ & $1.03^{+0.49}_{-0.25}$ \\
Norm & $1.67^{+0.57}_{-0.43}$ & $1.72^{+0.56}_{-0.44}$ & $7.31^{+2.08}_{-1.54}$ & $9.37^{+3.24}_{-2.44}$ \\
\hline
$E_{\rm line}$ & -- & $1.89^{+0.04}_{-0.03}$ & -- & $1.89^{+0.05}_{-0.04}$ \\
$\sigma$ & -- & $0.08^{+0.05}_{-0.03}$ & -- & $0.08^{+0.06}_{-0.08}$ \\
$S_{\rm line}$ & -- & $0.20^{+0.05}_{-0.03}$ & -- & $0.19^{+0.09}_{-0.08}$ \\
Sig. ($\sigma$)$^{a}$ & -- & $4.1$ & -- & $3.5$ \\
\hline
$F_X$ & $1.28^{+0.08}_{-0.05}$ & $1.24^{+0.06}_{-0.04}$ & $1.33^{+0.01}_{-0.01}$ & $1.27^{+0.01}_{-0.01}$ \\
$L_X$ & $1.52^{+0.07}_{-0.04}$ & $1.56^{+0.08}_{-0.06}$ & $1.58^{+0.01}_{-0.01}$ & $1.58^{+0.01}_{-0.01}$ \\
\hline
C/dof$^{b}$ & 74.9/59 & 53.9/54 & 72.0/56 & 54.8/53 \\
C/dof$^{c}$ & 63.07/42 & 43.03/39 & 61.70/41 & 44.35/38 \\
\hline
\end{tabular}
\tablefoot{Column labels denote the following models: A = \textit{tbabs*diskbb}, B = \textit{tbabs*gabs*diskbb}, C = \textit{tbabs*cutoffpl}, and D = \textit{tbabs*gabs*cutoffpl}. The \textit{diskbb} and \textit{cutoffpl} normalization is given in units of $10^{-2}$. The absorbed flux, $F_{\rm X}$, is given in units of $10^{-13}\ \mathrm{erg\ cm^{-2}\ s^{-1}}$, while the unabsorbed luminosity, $L_{\rm X}$, is given in units of $10^{39}\ \mathrm{erg\ s^{-1}}$, assuming a distance of 9.95 Mpc. $^{a}$: Monte Carlo significances are $4.1\sigma$ and $3.5\sigma$ for the \textit{tbabs*gabs*diskbb} and \textit{tbabs*gabs*cutoffpl} models. $^{b}$ and $^{c}$: C-statistics obtained from fits to spectra grouped to 10 and 15 counts per bin, respectively}.
\end{table}

We also examined the remaining four \textit{Chandra} observations and found that only ObsID 19497 ($\sim25$ ks) shows a hint of a similar feature (see Appendix~\ref{add}). The spectrum was adequately described by a blackbody (\textit{bbody}) model with $kT = 0.51 \pm 0.02$ keV, although the residuals showed a weak absorption-like feature near $\sim 1.9$ keV. Adding a \textit{gabs} component with $\sigma=0.08$ keV improved the fit by $\Delta C = 11.23$, yielding $E_{\rm line}=1.97\pm0.07$ keV and $S_{\rm line}=0.24^{+0.17}_{-0.13}$ keV (90\% confidence), while Monte Carlo simulations indicated a significance of only $\sim2.2\sigma$. This is most likely due to the limited photon statistics of the dataset: the net source count rate is only $\sim1.27\times10^{-2}$ counts s$^{-1}$ in the 0.3--10 keV band over an exposure of $\sim24.5$ ks.

\subsection{Timing}

We investigated the short-term variability of NGC\,4861 X--2 using background-subtracted light curves from the \textit{Chandra} observation ObsID 20992, in which the candidate absorption feature is most significantly detected. Additional details of the timing analysis are provided in Appendix~\ref{app:timing}. A $\chi^2$ test against a constant count-rate model reveals significant variability in the soft band (0.3--2.0 keV), with $\chi^2=92.81$ for 59 dof ($p=3.3\times10^{-3}$), whereas the hard band (2.0--10.0 keV) remains consistent with a constant flux, with $\chi^2=70.24$ for 58 dof ($p=0.13$) (Fig.~\ref{F:lcc3}). In addition, no statistically significant variability is detected in the hardness ratio (defined as the ratio of hard to soft count rates), indicating that the spectral shape remains stable on short timescales. This suggests that the observed variability is dominated by changes in the soft emission component.

We also searched for coherent periodic signals using the $Z_1^2$ (Rayleigh) test (see Appendix~\ref{app:timing} for details). A candidate periodic signal at $P = 7.4$ s is identified in the deepest \textit{Chandra} observation. The strongest peak is found in the soft X-ray band with $Z_1^2 = 28.32$ (see upper panel of Fig.~\ref{F:Z1}), corresponding to a single-trial significance of $\sim4.9\sigma$ and a global significance of $\sim2.5\sigma$ after accounting for the number of independent frequencies. The signal is also present in the full band (0.3--10 keV) with lower power ($Z_1^2 = 23.50$), corresponding to $4.5\sigma$ single-trial and $\sim1.5\sigma$ global significance. No significant peak is detected in the hard band. Folding the background-subtracted soft-band light curve (see lower panel of Fig.~\ref{F:Z1}) at this period, we obtain a pulse fraction of $41.4 \pm 10.5\%$, defined as $PF = (F_{\max}-F_{\min})/(F_{\max}+F_{\min})$. A consistent timing pattern is found in the shorter \textit{Chandra} observation (ObsID 19497; see Appendix~\ref{app:timing_c2} for details). A candidate signal at $P \approx 7.4$ s is detected with $Z_1^2 = 23.34$ (see Fig.~\ref{F:Z1_c2}) corresponding to a global significance of $\sim1.9\sigma$. The modulation is again confined to the soft X-ray band. The hardness ratio does not exhibit significant variations, indicating that the variability is driven by the soft emission component. No corresponding periodic signal is detected in the other available \textit{Chandra} observations.

\section{Discussion and conclusions}

We report the detection of an absorption feature at $\sim 1.89$ keV in the deepest 59 ks \textit{Chandra} observation of NGC 4861 X--2. The feature is consistently recovered across the adopted continuum models (\textit{diskbb} and \textit{cutoffpl}), with significant improvements in the fit statistic ($\Delta C \approx 17$--21 across the adopted continuum models and spectral grouping schemes). Monte Carlo simulations yield significances of $\sim3.5\sigma$ and $4.1\sigma$ for the \textit{cutoffpl} and \textit{diskbb} continua, respectively. To search for localized spectral features without prior assumptions on their energy, we performed a blind line scan over the 0.3--10 keV band (Fig.~\ref{fig:linescan}; see Appendix~\ref{blind}). These results indicate that the feature is localized and statistically robust.

 The inferred line width ($\sigma \sim 0.08$ keV; FWHM $\sim 0.19$ keV) exceeds the ACIS-S energy resolution at this energy (\citealp[$\sim$100--150 eV; FWHM]{2003SPIE.4851...89P}), suggesting that the measured width is broader than expected from the instrumental response alone. Although an atomic origin cannot be ruled out, as ULXs are known to host highly ionized outflows capable of producing narrow absorption features in their X-ray spectra (e.g., \citealt{2016Natur.533...64P,2021MNRAS.508.3569K}), a simple photoionized absorber scenario is disfavored since no accompanying transitions (e.g. Si XIV at $\sim$2.006 keV) are detected. A possible identification is the Si XIII He$\alpha$ complex at $\sim$1.86 keV, but the absence of expected additional lines weakens a straightforward atomic interpretation. Given the limited data quality, alternative scenarios such as absorption in outflowing material cannot be excluded. To further test whether the feature could be affected by oversampling near the instrumental Si K-edge, we regrouped the spectrum using the optimal binning method of \cite{2016A&A...587A.151K} (see Appendix A). The absorption-like feature near 1.89 keV remains visible with consistent best-fit parameters and a significance of $\sim3.6\sigma$ based on simulations. Finally, no significant gain shift is required, ruling out an instrumental origin for the feature.

An additional \textit{Chandra} observation (ObsID 19497) shows a weak ($\sim2.2\sigma$) feature at $1.97 \pm 0.07$ keV, consistent in energy with that detected in the deepest observation. The line centroid is consistent, within uncertainties, with that of the deepest \textit{Chandra} observation, indicating no significant energy shift between the two epochs. No statistically significant feature is detected in the other \textit{Chandra} or available \textit{XMM-Newton} observations, likely due to lower photon statistics, although intrinsic variability (e.g., in viewing geometry, accretion rate, or accretion-column structure; \citealt{2019A&A...622A..61S,2012A&A...544A.123B}) cannot be excluded. Moreover, we computed 90\% upper limits on the line strength for the remaining \textit{Chandra} observations without a significant feature by fixing the line energy and width to the values measured in ObsID 20992. The resulting upper limits are $S_{\rm line}<0.18$ keV, $<0.06$ keV, and $<0.06$ keV for ObsIDs 12473, 20993, and 21036, respectively. These limits are consistent with the line strength measured in the deepest observation, indicating that the non-detections can be explained by the lower data quality. The recurrence of the feature at the same energy in an independent observation disfavors a statistical fluctuation or a continuum modeling artifact. This behavior supports a common physical origin, most naturally interpreted as a CRSF, and implies a persistent and highly magnetized accretion environment.

The centroid energy of the feature at $\sim 1.89$ keV allows an estimate of the magnetic field strength under cyclotron interpretations. For a proton cyclotron origin, using $E_{\rm cyc,p}\simeq0.63\,B_{14}/(1+z)$ keV \citep{2006RPPh...69.2631H}, the observed energy implies $B \simeq (3.6$--$4.2)\times10^{14}$ G for gravitational redshifts $(1+z)=1.2$--1.4, implying magnetic field strengths in the magnetar field-strength range, while the source itself remains consistent with an accreting NS. In contrast, an electron cyclotron interpretation, based on $E_{\rm cyc,e}\simeq11.6\,B_{12}/(1+z)$ keV \citep{2019A&A...622A..61S}, yields $B \simeq (1.9$--$2.3)\times10^{11}$ G for the same redshift range. While such values are typical of accreting NSs, several models of ULX pulsars invoke strong magnetic fields to sustain super-Eddington accretion through magnetically channeled inflow onto the NS poles \citep{2015MNRAS.454.2539M,2017ARA&A..55..303K}. The inferred field likely reflects the local conditions in the line-forming region rather than the global dipole component, consistent with the presence of complex, possibly multipolar magnetic field geometries proposed for some ULX pulsars (e.g. \citealt{2017Sci...355..817I}). Proton cyclotron features have also been reported in other ULXs. In M51 ULX-8, an absorption feature at $\sim4.5$ keV has been interpreted as a proton CRSF \citep{2018NatAs...2..312B}, while a feature at $\sim3.3$ keV has been reported in NGC 4656 ULX-1 \citep{2026arXiv260310331C}. Compared to these sources, the lower energy of the feature in NGC 4861 X--2 implies a somewhat lower magnetic field strength, but still within the magnetar regime.

Assuming the absorption feature represents a proton cyclotron line, the observed width ($\sigma \approx 0.08$\,keV) likely requires additional broadening mechanisms beyond the instrumental response. In CRSFs, the line profile can be affected by pulse-phase averaging due to angle-dependent scattering and changing viewing geometry during the neutron-star rotation cycle. Therefore, part of the observed broadening may arise from phase-dependent variations of the line centroid and depth \citep{2013ApJ...771...96M}. However, if the measured width were interpreted purely as thermal Doppler broadening of a proton CRSF, the required plasma temperature would be unrealistically high ($kT \sim 1.5$ MeV). Similarly, rotational Doppler broadening alone would require spin frequencies much higher than the candidate $\sim7.4$ s periodicity inferred from the timing analysis. This suggests that magnetic-field gradients, geometric averaging, or scattering and reprocessing in optically thick funnel material may also contribute to the observed line width. One possible scenario is that an intrinsically narrower proton cyclotron feature becomes broadened during multiple scatterings in the funnel walls. In this case, the required electron temperature ($\sim0.9$ keV) is comparable to that inferred from the \textit{diskbb} continuum model.

The \textit{diskbb} model yields $kT_{\rm in}\sim 0.8$ keV and an apparent inner disk radius of $R_{\rm in}\sim130$ km (assuming a distance of 10 Mpc and a face-on geometry), corresponding to $\sim150$ km after standard corrections. Although larger than a typical NS radius, this estimate should not be interpreted literally, as the emission in ULXs is likely dominated by the accretion column and reprocessing in an optically thick wind or funnel. The \textit{cutoffpl} model instead indicates strong spectral curvature within the \textit{Chandra} band, with a low cutoff energy ($E_{\rm cut}\sim1.3$ keV) implying that the rollover occurs within the observed range. This suggests that the model primarily traces the intrinsic spectral curvature.

The candidate periodic signals detected in the two \textit{Chandra} observations show similar properties, with modulation confined to the soft X-ray band and no significant variability in the hard band. The centroid frequencies are consistent within uncertainties, with $f \approx 0.135\ \mathrm{Hz}$ (corresponding to $P \approx 7.4$ s) in both observations, although the limited statistical significance prevents a firm conclusion. Given the limited statistical significance and short exposure of ObsID 19497, we cannot distinguish between intrinsic spin evolution and orbital effects. The non-detection in the remaining datasets is consistent with limited photon statistics and possible frequency drift. The derived pulse fraction ($PF \sim 41--51\%$) is consistent with values typically observed in ULX pulsars, supporting the interpretation of an NS accretor.

In summary, we identify a localized absorption feature at $\sim 1.89$ keV that is robust against continuum modeling and consistent with a possible proton cyclotron origin. The observed line width may result from a combination of phase-dependent CRSF variability, magnetic-field gradients, and scattering or reprocessing in optically thick funnel material. A feature at the same energy is also tentatively detected in an additional \textit{Chandra} observation, providing further support for a physical origin. We also find candidate soft-band periodic signals at $P \approx 7.4$ s in two \textit{Chandra} observations, with properties consistent with a possible NS spin modulation. The fact that both the candidate periodic modulation and the absorption feature are detected in the same two observations may indicate a common physical origin, possibly related to the accretion column of a highly magnetized NS.

\begin{acknowledgements}
This work is partially supported by the Bundesministerium f\"ur Wirtschaft und Energie through the Deutsches Zentrum f\"ur Luft- und Raumfahrt e.V. (DLR) under the grant 50 OR 2517.
LD acknowledges funding from the Deutsche Forschungsgemeinschaft (DFG, German Research Foundation) - Projektnummer 549824807. VFS was supported by the DFG grant WE\,1312/59-1. W.Y. acknowledges support from the Alexander von Humboldt Foundation.
\end{acknowledgements}

\bibpunct{(}{)}{;}{a}{}{,}
\bibliographystyle{ngc4861ULX2}
\bibliography{ngc4861ULX2}

\begin{appendix}

\section{Energy spectra} \label{spec}

\begin{figure}
\resizebox{\hsize}{!}{\includegraphics{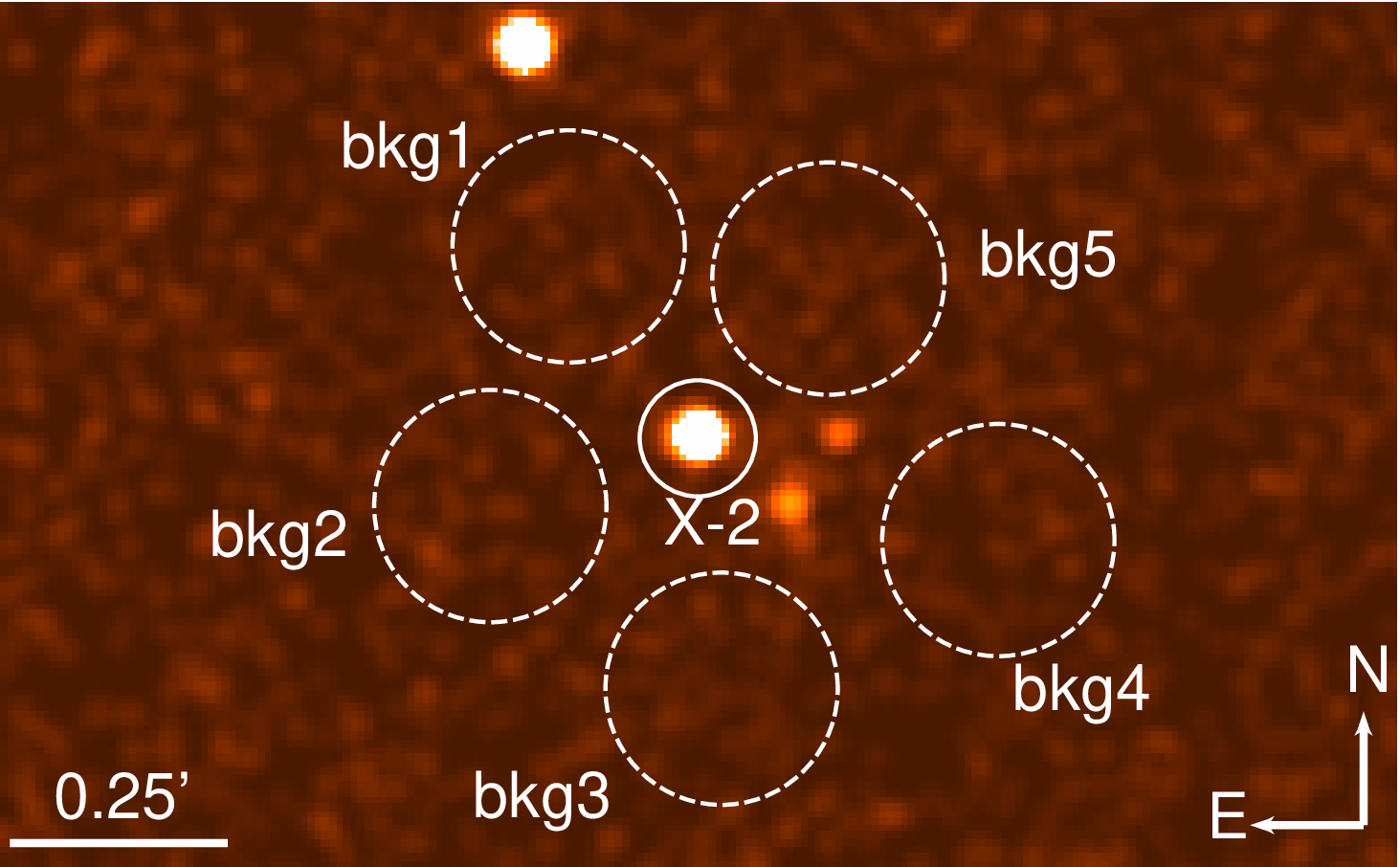}}
\caption{\textit{Chandra}/ACIS-S image (ObsID 20992) of NGC\,4861 X--2, showing the source extraction region (solid circle; radius $4^{\prime\prime}$) and five alternative background regions (dashed circles; radius $8^{\prime\prime}$) used to test the stability of the spectral results against background selection. The image is smoothed with a $3^{\prime\prime}$ Gaussian kernel.}
\label{F:bkg_regions}
\end{figure}

The \textit{Chandra} ACIS-S (Advanced CCD Imaging Spectrometer Spectroscopic array) observations were reduced using the \textit{Chandra} Interactive Analysis of Observations (CIAO; \cite{2006SPIE.6270E..1VF}) together with the most recent calibration database CALDB. Level 2 event files were generated using the standard \textit{chandra\_repro} pipeline. Spectra and light curves were extracted with the \textit{specextract} and \textit{dmextract} tools, respectively. For the \textit{Chandra} analysis, the source spectrum was extracted from a circular region with a radius of $4^{\prime\prime}$ centered on the source position. This radius encloses the majority of the \textit{Chandra}/ACIS point-spread function at the source position. Larger extraction regions were avoided due to the presence of nearby sources, which could lead to contamination. Background spectra were extracted from nearby source-free circular regions with radii of $8^{\prime\prime}$ on the same CCD. In addition, five independent background regions were tested (see Fig.~\ref{F:bkg_regions}) to assess whether the spectral results depend on the choice of background. Background spectra extracted from these regions are consistent with each other and do not show any feature at $\sim 1.9$ keV. In addition, the pile-up fraction was estimated using the CIAO task \textit{pileup\_map}, yielding a maximum value of $<3\%$ within the source extraction region. This level is well below that at which pile-up is expected to significantly distort the spectral shape in ACIS observations, and therefore, pile-up does not affect our spectral results.

The \textit{XMM-Newton} EPIC data were reduced using the Science Analysis System (\textit{SAS}) v22.0 with the most recent calibration files. Calibrated event lists were produced using \textit{epproc} for EPIC-pn and \textit{emproc} for EPIC-MOS. Periods of high particle background were filtered out using full-field high-energy light curves to define good-time intervals. Valid X-ray events were selected by requiring PATTERN$\leq$4 for pn and PATTERN$\leq$12 for MOS, together with \textit{FLAG==0}. However, the \textit{XMM--Newton} analysis is affected by contamination from nearby variable X-ray sources. We adopted a relatively small extraction region to minimize this effect; however, two nearby variable X-ray sources lie within $\sim8^{\prime\prime}$ of X--2, and their contribution cannot be fully excluded given the EPIC PSF. We also tested smaller extraction regions, but no robust result was obtained. We therefore do not use the \textit{XMM--Newton} data to constrain the presence of the $\sim 1.9$ keV feature.

\begin{figure}
\resizebox{\hsize}{!}{\includegraphics{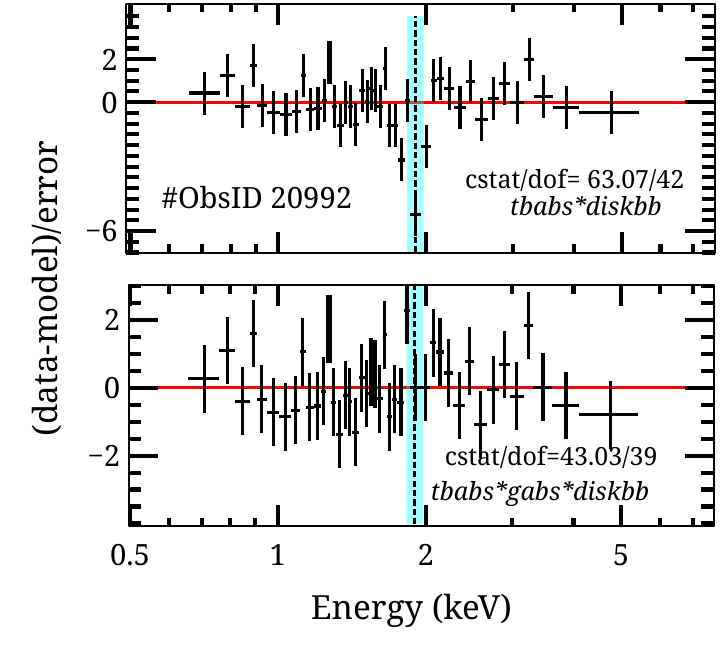}}
\caption{Residuals of the \textit{Chandra}/ACIS-S spectrum (ObsID 20992) of NGC\,4861 X--2, fitted using C-statistics (cstat) after grouping 15 counts per bin. The top panel shows the residuals for the continuum-only \textit{tbabs*diskbb} model, while the bottom panel shows the residuals for the \textit{tbabs*gabs*diskbb} model. The shaded region highlights the absorption-like feature at $\sim 1.89$ keV.}
\label{fig:residual_grouping15cstat}
\end{figure}

\textit{Chandra} spectra were grouped to ensure a minimum of 10 and 15 counts per bin, providing a balance between spectral resolution and statistical reliability, and all spectral fits were performed using C-statistics (cstat). The resulting best-fit parameters obtained with the different grouping schemes are consistent within uncertainties. Parameter errors were estimated at the 90\% confidence level using the XSPEC \textit{steppar} command. To assess the robustness of the detected feature against spectral grouping, spectra grouped with minimum bin sizes of 10 and 15 counts per bin were independently examined using C-statistics, and both grouping schemes consistently reveal the absorption feature at $\sim$1.89 keV (Fig.~\ref{fig:residual_grouping15cstat}). For the \textit{Chandra} spectra, the intrinsic absorption is generally low and not well constrained. We therefore fixed $N_{\rm H}$ to the Galactic value, $N_{\rm H}=1.3\times10^{20}\ \mathrm{cm^{-2}}$, for continuum models in which no additional absorption was required by the data. To exclude any instrumental or background-related origin, background spectra were extracted from multiple source-free regions on the same CCD and analyzed independently, yielding no absorption features near $\sim 1.9$ keV. We focus on the deepest \textit{Chandra}/ACIS-S observation (ObsID 20992; $\sim$58 ks), which provides the highest-quality spectrum. The \textit{gabs*diskbb} and \textit{gabs*cutoffpl} models provide statistically and physically consistent descriptions of the spectrum. In both cases, the inclusion of the Gaussian absorption component yields a significant improvement in the fit, with continuum parameters remaining consistent within uncertainties.

The \textit{power-law} model was also tested as a single-component continuum; however, it does not provide an adequate description of the spectrum. The fit converges to a steep photon index ($\Gamma \sim 2.7$) accompanied by increased intrinsic absorption, reflecting a degeneracy between $\Gamma$ and $N_{\rm H}$ due to the lack of intrinsic curvature in the model. Consequently, significant structured residuals remain across the 0.5--8 keV band. Nevertheless, a negative residual is still detected at $\sim1.88$ keV, and the addition of a Gaussian absorption component improves the fit by
$\Delta C \approx 15$, corresponding to a significance of $\sim2.7\sigma$ based on Monte Carlo simulations. However, given the poor representation of the continuum, this result is not considered robust, and the \textit{power-law} model is not included in the main analysis.

\begin{table}
\caption{Summary of the observations used in this work.}
\label{tab:obs_log}
\centering
\begin{tabular}{cccc}
\hline\hline
\multicolumn{4}{c}{\textit{Chandra}} \\
\hline
Obs ID & Instr. & Exp. (ks) & Start Date \\
\hline
12473 & ACIS-I	 & 19.78 & 2012-01-03 \\
19497 & ACIS-S & 24.52 & 2018-03-07 \\
20992 & ACIS-S & 58.46 & 2018-03-11 \\
20993 & ACIS-S & 27.69 & 2018-03-16 \\
21036 & ACIS-S & 37.78 & 2018-03-16 \\
\hline
\multicolumn{4}{c}{\textit{XMM-Newton}} \\

\hline
0141150101 & EPIC & 28.51 & 2003-06-14 \\
0141150401 & EPIC & 14.52 & 2003-07-10 \\
0141150501 & EPIC & 21.85 & 2003-12-03 \\
\hline
\end{tabular}
\end{table}

\begin{figure}
\resizebox{\hsize}{!}{\includegraphics{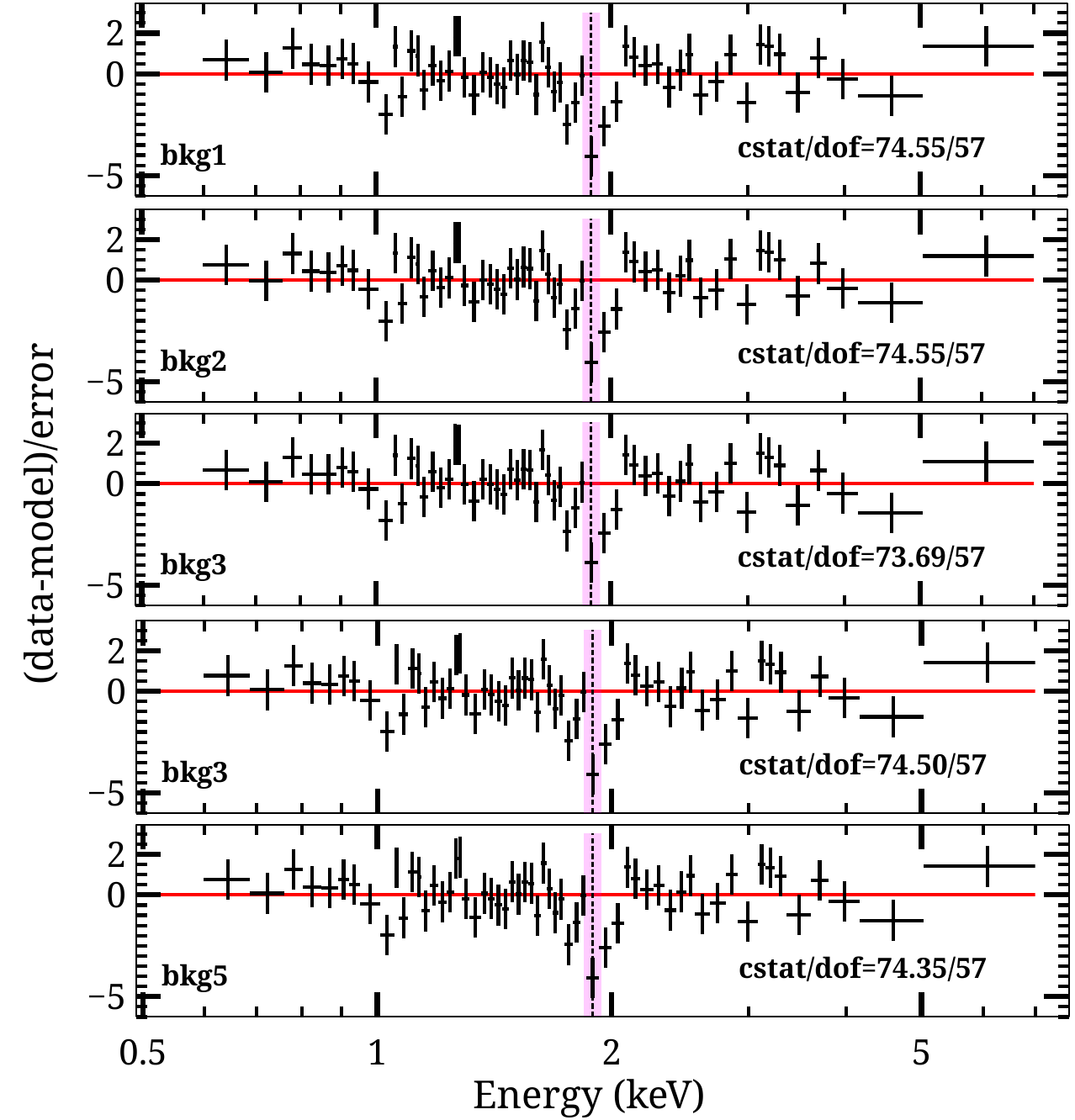}}
\caption{Residuals of the \textit{Chandra}/ACIS-S source spectrum (ObsID 20992) of NGC\,4861 X--2 obtained using five independent background regions (bkg1--bkg5; see Fig.~\ref{F:bkg_regions}). In all cases, the source spectrum was fitted with the same continuum model, \textit{tbabs*diskbb}. The residuals remain consistent across different background selections, and the absorption-like feature at $\sim 1.89$ keV persists in all cases.}
\label{fig:background_test}
\end{figure}

\subsection{Blind line scan and Monte Carlo simulations} \label{blind}

For each continuum model, $10^5$ spectra were simulated under the null hypothesis, that is, without the \textit{gabs} component, and the likelihood-ratio improvement was compared with that obtained from the real data. In each simulation, the line centroid energy was allowed to vary over the searched energy range, following the same fitting procedure applied to the real spectrum. For the \textit{diskbb} continuum, only 2 out of $10^5$ simulations exceeded the observed likelihood ratio, corresponding to a $p$-value of $2.0 \times 10^{-5}$ ($4.1\sigma$). For the \textit{cutoffpl} continuum, we obtain $p = 4.3 \times 10^{-4}$ ($3.5\sigma$).

To search for localized spectral features without prior assumptions on their energy, we performed a blind line scan over the 0.3--10 keV range using XSPEC. Starting from the best-fit continuum-only model, a multiplicative Gaussian line component (\textit{gabs}) was added, and its centroid energy was stepped across the spectrum on a uniform grid with a step size of 0.02 keV. At each trial energy, the line centroid was fixed, while the line strength was allowed to vary, and the line width was fixed to $\sigma = 0.08$ keV, consistent with the best-fit value obtained from the spectral analysis. The fit was then re-optimized, and the improvement in fit statistic relative to the continuum-only model was computed as $\Delta C$. The spectra were optimally grouped following the \cite{2016A&A...587A.151K} prescription implemented in \texttt{ftgrouppha}, and fitted using C-statistics. All continuum parameters were allowed to vary freely at each step. The scan was performed for different continuum models, including \textit{tbabs*diskbb} and \textit{tbabs*cutoffpl}, to test the robustness of the results against continuum uncertainties.

In both cases, a single localized peak is detected at $\sim1.88$--$1.90$ keV. The maximum improvement reaches $\Delta C \approx 19$ for the \textit{diskbb} continuum and $\Delta C \approx 15$ for the \textit{cutoffpl} model. No other features with comparable $\Delta C$ values are found across the scanned energy range. The adopted energy range and step size adequately sample the spectral resolution of \textit{Chandra}/ACIS, and the consistency of the detected feature across different continuum models indicates that it is robust against continuum uncertainties. In addition, a weak excess is present near $\sim 0.95$ keV in the \textit{cutoffpl} blind scan, approximately half of the main feature energy. However, its $\Delta C$ is very small compared to the dominant peak at $\sim 1.9$ keV, and we therefore do not consider it statistically significant.

\begin{figure} \resizebox{\hsize}{!}{\includegraphics{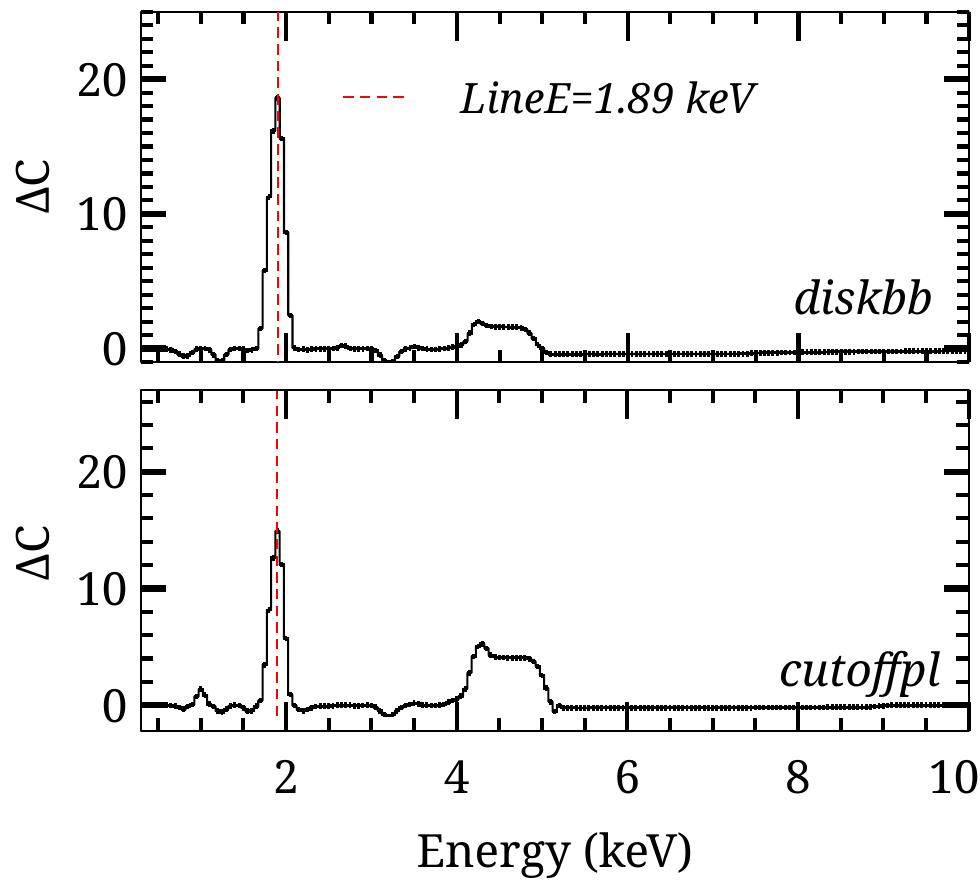}} 
\caption{Blind line scan of the optimally binned \textit{Chandra}/ACIS-S (ObsID 20992) spectrum of NGC\,4861 X--2, performed over the 0.3--10 keV energy range using C-statistics. \textit{Top panel:} improvement in fit statistic ($\Delta C$) as a function of trial line energy obtained with the \textit{tbabs*gabs*diskbb} model relative to the continuum-only \textit{tbabs*diskbb} model. \textit{Bottom panel:} corresponding results for the \textit{tbabs*gabs*cutoffpl} model relative to the continuum-only \textit{tbabs*cutoffpl} model. A notable improvement is consistently found near $\sim$1.89 keV in both continuum models.}
\label{fig:linescan} 
\end{figure}

\subsection{Evidence for the 1.9 keV feature in an additional \textit{Chandra} observation} \label{add}

\begin{figure}
\resizebox{\hsize}{!}{\includegraphics{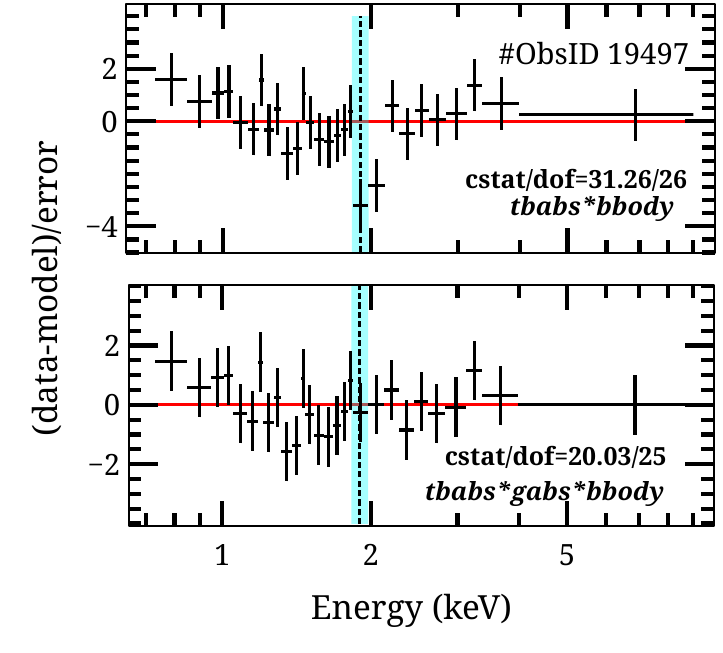}}
\caption{Residuals of the \textit{Chandra}/ACIS-S spectrum (ObsID 19497) fitted with a blackbody continuum model. \textit{Top panel:} Residuals for the \textit{tbabs*bbody} model, showing a negative deviation around $\sim 1.9$ keV (shaded region). \textit{Bottom panel:} Residuals after adding a Gaussian absorption component (\textit{tbabs*gabs*bbody}), where the feature is largely removed. The vertical dashed line marks the centroid energy of the feature.}
\label{F:add}
\end{figure}

We examined all available \textit{Chandra} and \textit{XMM-Newton} observations of NGC\,4861 X--2 to search for the presence of a similar absorption feature. Among these, only the \textit{Chandra}/ACIS-S observation ObsID 19497 (exposure $\sim25$ ks) shows a hint of a feature consistent with that detected in the deepest observation (ObsID 20992) (see Fig.~\ref{F:add}), while no significant feature is found in the remaining observations. The ObsID 19497 spectrum was grouped to 10 counts per bin, resulting in 28 spectral bins, and fitted in the 0.3--10 keV band using C-statistics. We first modeled the continuum with an absorbed blackbody model (\textit{tbabs*bbody}), obtaining $C = 31.26$. The best-fit temperature is $kT = 0.51 \pm 0.02$ keV, while the absorption column density was fixed to the Galactic value. The residuals show a weak absorption-like structure near $\sim 1.9$ keV. The unabsorbed luminosity in the 0.3--10 keV band is $L_{\rm X} \approx 1.23 \times 10^{39}\ {\rm erg\ s^{-1}}$. We also tested the \textit{diskbb} and \textit{cutoffpl} models for ObsID 19497; however, due to limited photon statistics, these models do not provide additional constraints compared to the \textit{bbody} model.

To test this feature, we added a multiplicative Gaussian absorption component (\textit{gabs}). Fixing the line width to the value measured in the deepest \textit{Chandra} observation ($\sigma = 0.08$ keV) while allowing the centroid energy to vary freely yields a best-fit energy of $E_{\rm line}=1.97^{+0.07}_{-0.07}$ keV and a line strength of $S_{\rm line}=0.24^{+0.17}_{-0.13}$ keV (90\% confidence). The fit improves to $C = 20.03$, corresponding to $\Delta C = 11.23$. The line energy is therefore consistent, within uncertainties, with the $\sim 1.89$ keV feature detected in the deepest observation.
As an additional consistency check, we fixed both the line energy and width to the values obtained from the deepest data ($E_{\rm line}=1.89$ keV and $\sigma=0.08$ keV), allowing only the line strength to vary. In this case, the fit improves to $C = 23.06$ ($\Delta C = 8.20$), with $S_{\rm line}=0.17^{+0.13}_{-0.11}$ keV. The lower bound of the line strength remains above zero at the 90\% confidence level, indicating that a feature consistent with that detected in the deepest observation is present in the data. To further assess the feature significance, we performed simulations with the XSPEC \textit{simftest} routine using $10^4$ realizations for the model with free centroid. The resulting significance is only $\sim2.2\sigma$, indicating that the feature is not formally significant in this observation. 

\subsection{Instrumental and background checks} \label{Silikon}

We performed several tests to assess a possible instrumental or background-related origin of the feature. First, background spectra extracted from multiple independent source-free regions on the same CCD do not show any feature near $\sim 1.9$ keV. Second, the absorption feature persists under different background selections and spectral groupings, indicating that it is not sensitive to data reduction choices. Third, allowing for a linear gain shift in the spectral fits does not improve the fit and does not remove the feature, ruling out a calibration-related energy offset. Finally, the feature is independently recovered across different continuum models and is also detected at a consistent energy in an additional \textit{Chandra} observation. Taken together, these tests strongly disfavor an instrumental or background-related origin and support a physical interpretation of the feature.

\begin{figure}
\resizebox{\hsize}{!}{\includegraphics{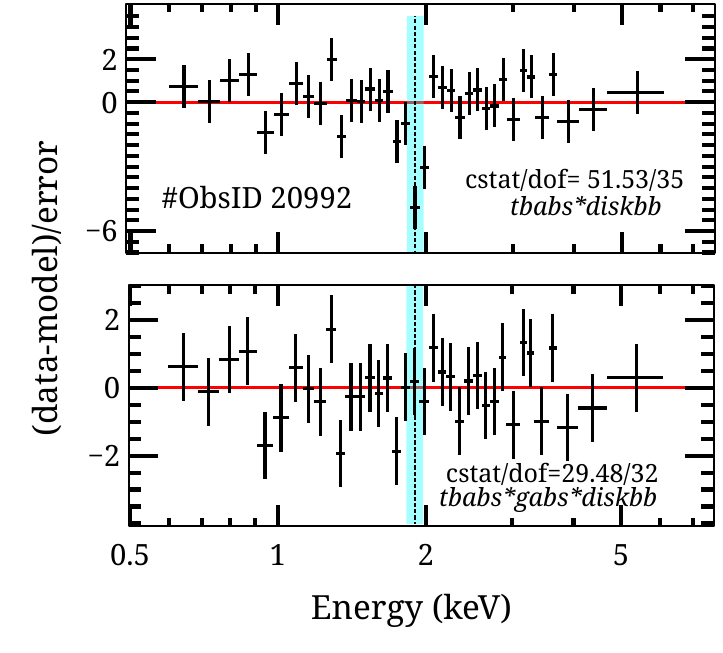}}
\caption{Top panel: Chandra/ACIS spectrum (ObsID 20992) of NGC\,4861 X--2 fitted with \textit{tbabs*diskbb} using C-statistics and optimally grouped following the \cite{2016A&A...587A.151K} method implemented with \textit{ftgrouppha}. Pronounced negative residuals are visible near $\sim$1.89 keV. Bottom panel: Residuals obtained after including \textit{gabs} in the model (\textit{tbabs*gabs*diskbb}). The absorption-like residuals around $\sim$1.89 keV are substantially reduced after the inclusion of the absorption component.}
\label{fig:residual_newbin}
\end{figure}

To further investigate whether the absorption-like feature near 1.89 keV could be influenced by oversampling effects close to the instrumental Si K-edge, we regrouped the Chandra/ACIS-S spectrum using the optimal binning method of \cite{2016A&A...587A.151K}, implemented through the \textit{HEASOFT} task \textit{ftgrouppha}. The spectrum was grouped using the \textit{optmin} method together with the corresponding response matrix, resulting in a grouping that avoids oversampling of the instrumental energy resolution while preserving the available spectral information. The spectral analysis was performed using C-statistics. We repeated the spectral fitting using both the \texttt{tbabs*diskbb} and \textit{tbabs*gabs*diskbb} models. The absorption-like feature near 1.89 keV remains visible in the residuals and is recovered with consistent best-fit parameters compared to the standard grouping adopted in the main analysis. Figure~\ref{fig:residual_newbin} shows the residuals obtained with the \textit{tbabs*diskbb} and \textit{tbabs*gabs*diskbb} models for the optimally grouped spectrum. We additionally repeated the Monte Carlo simulations for the \textit{tbabs*gabs*diskbb} model using the optimally grouped spectrum. Based on 10,000 simulations, the feature remains significant at the $\sim 3.6\sigma$ confidence level. The Monte Carlo simulations were performed by allowing the line energy to vary within the investigated energy range, following the same fitting procedure adopted for the real spectrum. These results suggest that the detected feature is unlikely to be produced by oversampling effects or statistical fluctuations associated with the instrumental Si K-edge region.

\section{Timing}\label{app:timing}

\begin{figure}
\resizebox{\hsize}{!}{\includegraphics{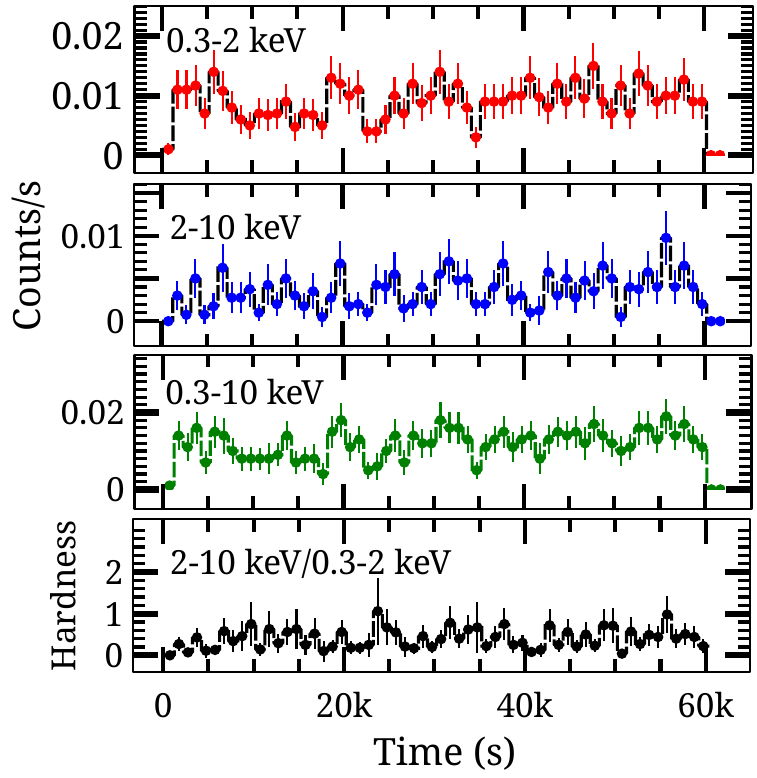}}
\caption{Background-subtracted \textit{Chandra}/ACIS-S (ObsID 20992) light curves of NGC\,4861 X--2. From top to bottom: soft band (0.3--2 keV), hard band (2--10 keV), total band (0.3--10 keV), and hardness ratio defined as the ratio of hard to soft count rates. The light curves are binned at 1000 s.}
\label{F:lcc3}
\end{figure}

Background-subtracted light curves were extracted from all available \textit{Chandra} and \textit{XMM-Newton} observations in the soft (0.3--2 keV) and hard (2--10 keV) energy ranges. For \textit{Chandra}, the CIAO tool \textit{dmextract} was used, while for \textit{XMM-Newton} we employed the SAS task \textit{evselect}. All event times were corrected to the solar system barycenter using \textit{axbary} and \textit{barycen}. Time-resolved hardness ratios were constructed as the ratio of hard (2--10 keV) to soft (0.3--2 keV) count rates in order to probe possible spectral variability. However, the hardness ratio does not show statistically significant variations within the uncertainties, indicating that the overall spectral shape remains broadly stable on short timescales. To quantify the short-term variability, we performed a $\chi^2$ test against a constant count-rate model. A threshold of $p < 0.01$ was adopted to claim significant variability. The source is significantly variable in the soft band ($\chi^2=92.81$ for 59 dof; $p=3.3\times10^{-3}$), whereas the hard band does not show statistically significant variability ($\chi^2=70.24$ for 58 dof; $p=0.13$). This indicates that the short-term variability is dominated by the soft X-ray emission, while the hard component remains comparatively stable (see Fig.~\ref{F:lcc3}).

We searched for coherent periodic signals using the $Z_1^2$ (Rayleigh) test in the full (0.3--10 keV), soft (0.3--2 keV), and hard (2--10 keV) energy bands. In the full band, the strongest peak is detected at $f = 0.13505$ Hz ($P = 7.4$ s) with $Z_1^2 = 23.50$, corresponding to a single-trial probability of $7.9\times10^{-6}$ ($4.47\sigma$). After accounting for the number of independent frequencies, the significance decreases to $1.5\sigma$, indicating that the signal is not statistically significant. In the soft band, the same periodicity is recovered at $P = 7.4$ s with higher power ($Z_1^2 = 28.32$), yielding a single-trial probability of $ \approx 10^{-6}$ ($4.9\sigma$), which reduces to a global significance of $2.5\sigma$ after trial correction. Although below the formal detection threshold, the enhancement in the soft band suggests that any modulation is primarily associated with the soft X-ray emission. 

No significant peak was detected in the hard band, and thus, no meaningful constraint on periodic variability can be obtained in this energy range. Likewise, no significant signal was found in the power density spectrum, likely due to limited photon statistics and the time resolution of the detector. Folding the soft-band light curve at $P = 7.4$ s yields a pulse fraction of $PF = 41.4 \pm 10.5$\%, where $PF = (F_{\max}-F_{\min})/(F_{\max}+F_{\min})$, which should be regarded as indicative given the limited statistical significance of the signal.

\begin{figure}
\resizebox{\hsize}{!}{\includegraphics{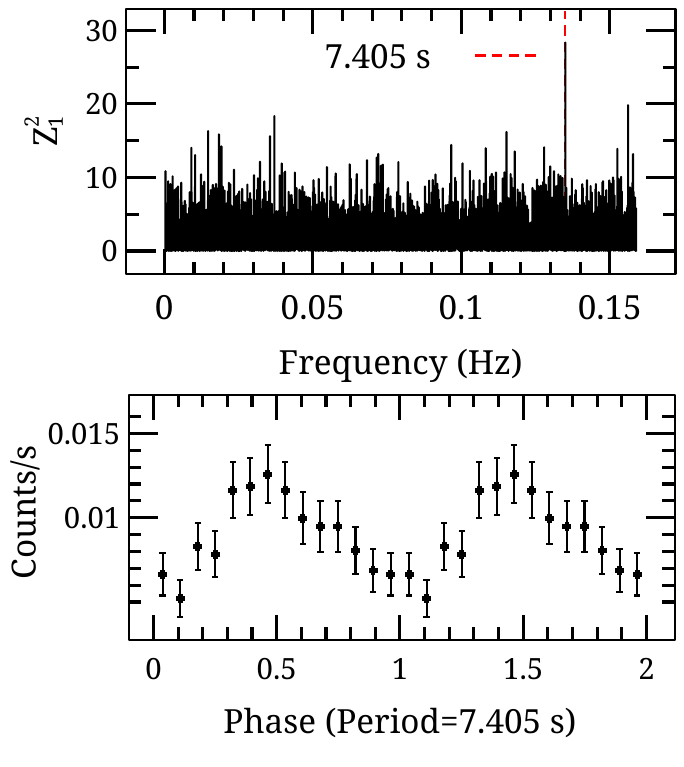}}
\caption{Timing analysis of NGC\,4861 X--2 in the soft X-ray band (0.3--2 keV) (ObsID 20992). \textit{Upper panel:} $Z_1^2$ periodogram as a function of trial frequency. The dashed red line indicates the position of the strongest peak at $f = 0.135$ Hz ($P = 7.4$ s). \textit{Bottom panel:} Folding the background-subtracted light curve at $P = 7.4$ s, shown over two cycles for clarity. Although the statistical significance is limited, the folded profile suggests a sinusoidal modulation with a candidate pulse fraction of $\sim41\%$.}
\label{F:Z1}
\end{figure}

\subsection{Additional timing analysis of ObsID 19497} \label{app:timing_c2}

\begin{figure}
\resizebox{\hsize}{!}{\includegraphics{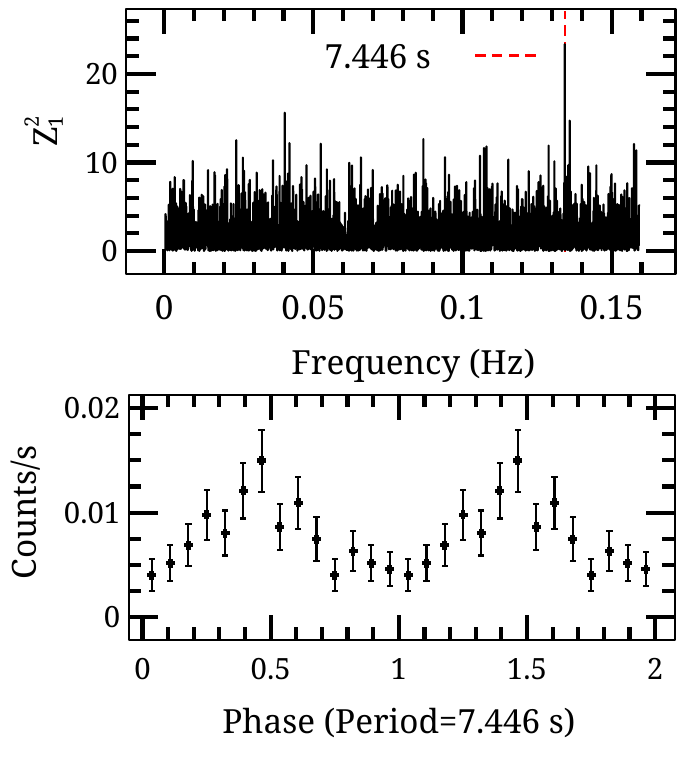}}
\caption{\textit{Chandra}/ACIS-S (ObsID 19497) soft-band (0.3--2 keV) timing analysis. \textit{Top panel:} $Z_1^2$ periodogram as a function of trial frequency, showing a peak at $P \approx 7.4$ s (dashed red line). \textit{Bottom panel:} background-subtracted light curve folded at $P = 7.4$ s, shown over two cycles for clarity.}
\label{F:Z1_c2}
\end{figure}

We performed a timing analysis of the \textit{Chandra}/ACIS-S observation ObsID 19497 (exposure $\sim24.5$ ks) following the same procedure applied to the deepest observation. Background-subtracted light curves were extracted in the soft (0.3--2 keV) and hard (2--10 keV) energy bands. A $\chi^2$ test against a constant count-rate model indicates variability in the soft band, while the hard band remains consistent with a constant flux. The hardness ratio, defined as the ratio of hard to soft count rates, remains below $\sim 0.8$ and does not show statistically significant variations, suggesting that the spectral shape is stable on short timescales. We searched for periodic signals using the $Z_1^2$ (Rayleigh) test. The strongest peak is detected at $f = 0.134300$ Hz ($P \approx 7.4$ s) with $Z_1^2 = 23.34$. This corresponds to a single-trial probability of $1.6\times10^{-5}$ ($\sim4.3\sigma$), which decreases to $\sim1.85\sigma$ after correcting for the number of independent trials. The signal is primarily driven by the soft band, while no significant peak is detected in the hard band. The $Z_1^2$ statistic shows no significant enhancement when including higher harmonics, indicating that the signal is consistent with a sinusoidal modulation (see Fig.~\ref{F:Z1_c2}). Folding the soft-band light curve at $P \approx 7.4$ s yields a pulse fraction of $PF = 51.6 \pm 14.8\%$. Given the limited photon statistics, this value should be regarded as indicative.

\end{appendix}

\end{document}